\documentclass[10pt]{article}

\usepackage{amssymb,amsthm,amsmath,amsfonts}

\usepackage{dcolumn}
\usepackage{rotating}

\usepackage{color}

\pdfoutput=1
\usepackage{graphicx}
\usepackage{xcolor}

\newcommand{\ba}{\begin{eqnarray}}
\newcommand{\ea}{\end{eqnarray}}
\newcommand{\be}{\begin{equation}}
\newcommand{\ee}{\end{equation}}
\newcommand{\bi}{\begin{itemize}}
\newcommand{\ei}{\end{itemize}}

\def\beq{\begin{equation}}
\def\eeq{\end{equation}}

\def\Fc{\mathcal{F}}

\def\Oc{\mathcal{O}}

\def\Rc{\mathcal{R}}

\newcommand{\const}{\text{const}}
\title{Post-inflationary GW production in generic higher (infinite) derivative gravity\thanks{Preprint: Imperial/TP/2022/AAT/2}}

\author{Alexey S. Koshelev$^{1}$, Alexei A. Starobinsky$^{2,3}$ and Anna Tokareva$^{4}$
\\ \mbox{}$^{1}$\textit{\small Departamento de F\'isica, Centro de Matem\'atica e Aplica\c{c}\~oes (CMA-UBI),}
\\\textit{\small Universidade da Beira Interior, 6200 Covilh\~a, Portugal}
\\ \mbox{}$^{2}$\textit{\small L. D. Landau Institute for Theoretical Physics RAS, Chernogolovka, Moscow region 142432, Russian Federation}
\\ \mbox{}$^{3}$\textit{\small Kazan Federal University, Kazan 420008, Republic of Tatarstan, Russian Federation}
\\ \mbox{}$^{4}$\textit{\small Theoretical Physics, Blackett Laboratory, Imperial College London, SW7 2AZ London, U.K.}
}

\date{~}

\begin{document}
\maketitle
\begin{abstract}
Gravity can be embedded into a renormalizable theory by means of adding quadratic in curvature terms. However, this at first leads to the presence of the Weyl ghost. It is possible to get rid of this ghost if the locality assumption is weakened and the propagator of the graviton is represented by an entire function of the d'Alembertian operator without new poles and zeros. Models of this type admit a cosmological solution describing the $R^2$, or Starobinsky, inflation. We study graviton production after inflation in this model and show that it is negligible despite the presence of the higher derivative operators which could potentially cause instabilities.
\end{abstract}

\section{Introduction}

Incorporating the very much successful purely geometric inflationary $R^2$ (more exactly, $R+R^2$) model where $R$ is the Ricci scalar \cite{Starobinsky:1980te,Starobinsky:1981vz,Starobinsky:1983zz} in a UV complete quantum theory is the goal many try to achieve. Construction of quantum gravity hits the problem of unitarity like it has happened for the model by Stelle \cite{Stelle:1976gc,Stelle:1977ry}. That model is based on the general quadratic in curvatures local action which has to include the Weyl tensor squared term leading to the presence of the Weyl ghost. On the other hand, the $R^2$ (Starobinsky) inflationary model does not contain ghosts, though it contains a massive scalar degree of freedom (dubbed scalaron in~\cite{Starobinsky:1980te}) which plays the role of inflaton in this model. It is also crucial for generation of primordial scalar perturbations in the Universe from quantum vacuum fluctuations during inflation~\cite{Mukhanov:1981xt,Starobinsky:1983zz}.

It was shown in a fairly novel attempt to understand the gravity theory better \cite{Efimov:1987opr,Tomboulis:1997gg,Biswas:2005qr,Modesto:2011kw} that analytic infinite derivative (AID) generalization may prove useful and essential in the construction of a UV complete theory. In particular, the result of \cite{Biswas:2016etb,Biswas:2016egy} shows that a generic diffeomorphism invariant gravity theory with analytic dependence on curvatures and covariant derivatives would lead to a higher derivative propagator but only \textit{infinite} number of derivatives can ultimately eradicate ghosts.
Concurring with this statements fundamental approaches like strings or higher spin theories manifestly lead to infinite derivative models. Starobinsky inflation was successfully embedded in the AID framework being its exact particular solution. A number of observationally crucial results were obtained regarding the power spectra, corrections to the tensor-to-scalar ratio and to the scalar non-gaussianities.

In this letter we consider the question of the graviton production as the result of the scalaron decay after inflation in this model. Although naively higher derivatives may lead to more than needed gravitons, our analysis shows that it is not the case.
As an essential step, we start by observing that tensor and scalar sectors in the model can be tied even further compared to the previous studies as long as the absence of ghosts is analyzed by invoking the ADM formalism. This would not lead to any immediate drastic changes in the previously obtained results but will be shown to be important for the suppression of production of gravitons compared to all other particles and antiparticles during creation and (re)heating of matter after inflation in the course of the scalaron decay.


\section{The model and the revisited ghost-free condition}

We start with the Analytic Infinite Derivative (AID) action which was intensively studied in \cite{Koshelev:2016xqb,Koshelev:2017tvv} and which reads
\begin{equation}
    S=\int d^4x\sqrt{-g}\left(\frac{M_P^2}2R+\frac\lambda 2\left(R\Fc_R(\square)R+W\Fc_W(\square)W\right)\right)
    \label{model}
\end{equation}
where $W$ is the Weyl tensor for which we almost always omit its index structure as it is irrelevant for what follows, $M_P$ is the reduced Planck mass, and we assume that form-factors $\Fc_{R,W}$ have Taylor expansions at zero arguments such that
\begin{equation}
    \Fc_{R,W}(\square)=\sum_{n\geq0} {f_{R,W}}_n\square^n \, .
\end{equation}
Existence of the Taylor series in the origin for the form-factors means that there is a well-defined IR limit.

Working in the ADM formalism and using the symmetry decomposition with respect to the 3-dimensional symmetry group, we can show that the kinetic term in the tensor sector on top of Minkowski space-time has the form
\begin{equation}
    h_{ij}^{TT}\left(1+2\square \Fc_{W}(\square)/M_P^2\right)\square h_{ij}^{TT}
\end{equation}
where $TT$ stays for transverse and traceless. In the scalar sector, we obtain for the canonically normalized scalar field
\begin{equation}
    \varphi\hat{O}_2\varphi=M_P^2\varphi\frac{(6\square \Fc(\square)/M_P^2-1)(1+2\square \Fc_{W}(\square)/M_P^2)}{2(\Fc(\square)+\Fc_W(\square)/3)}\varphi \, .
\end{equation}
We impose the ghost-free conditions for both operators by requiring that there are no poles in the propagators apart from those which exist in $f(R)$ modified gravity which is a particular case of scalar-tensor gravity. It is not that we just do not want any new modes,  but rather that such modes will be inevitably ghosts following the long-standing Ostrogradski statement \cite{ostro}.

In the tensor sector this leads to
\begin{equation}
    \Fc_W(\square)=M_P^2\frac{e^{\sigma(\square)}-1}{2\square}
\label{p2}
\end{equation}
with $\sigma$ being an entire function. Mathematically exponent of an entire function is the only construction which has no zeros on the whole complex plane. We notice that the Stelle gravity case corresponds to $\Fc_W=\const$ and leads to a second pole, the Weyl ghost. Moreover, it is obvious that in the presented construction no finite number of derivatives can help exorcising the ghost.

In the scalar sector we expect to have a scalaron with the mass $M$ meaning that
\begin{equation}
\hat{O}_2=e^{\sigma_1(\square)}(\square-M^2)\, .
\label{p0}
\end{equation}
Again $\sigma_1$ must be an entire function. Our essential assumption is that $M$ is significantly less than $M_P$, so that we can safely choose the non-locality scale $\Lambda$ to be much more than $M$, but still much less than $M_P$. It is justified by the fact that in the local $R^2$ model, $M^2=\frac{24\pi^2\cal{P}_{{\cal R}}}{N^2}M_P^2$ where $N$ is the number of e-folds from the end of inflation and ${\cal P}_{{\cal R}}$ is the power spectrum of primordial scalar perturbations generated during inflation~\cite{Starobinsky:1983zz}\footnote{Note that in this model, ${\cal P}_{{\cal R}}\propto N^2$ for $N\gg 1$, so $M$ does not depend on $N$.}. Using the observational result ${\cal P}_{{\cal R}}\approx 2.1\cdot 10^{-9}$ at the pivot scale $k=k_{\ast}=0.05$ Mpc$^{-1}$~\cite{Planck:2018jri}, we get $M\approx 1.3\cdot 10^{-5} M_P$ for $N(k_{\ast})=55$. In fact, $N(k_{\ast})$ can be in the range $(50-60)$ depending on the most effective channel of scalaron decay into known particles (e.g., into minimally or non-minimally coupled light scalars, or massive heavy fermions) after the end of inflation. Thus, the smallness of $M/M_P$ is due to the smallness of ${\cal P}_{{\cal R}}(k_{\ast})$ which can be considered as a new fundamental cosmological constant. So, our assumption is that $M$ remains practically the same in our non-local gravity model if $\Lambda\gg M$, and our calculations confirm it. 

Two comments are in order here. The first is that the above expressions are quite different from those obtained for example in \cite{Biswas:2016egy}. This should not be confusing because in that other paper 4-dimensional symmetry spin decomposition was used, and as such what was named transverse and traceless mode there is not yet the physical graviton with 2 polarizations. It rather mixes with physical scalar sector and as such one still has to resolve constraint equations. The second comment is that in contrary to the local $R^2$ gravity (or generic $f(R)$ gravity), we can in principle demand that the scalar mode does not show up in the spectrum at all. This can be done if we require $\hat{O}_2=e^{\sigma_1(\square)}$, which is not the case of interest though since it does not provide us a possibility to generate primordial matter perturbations in the Universe from quantum vacuum fluctuations during inflation.

The requirement (\ref{p0}) translates into
\begin{equation}
    \Fc_R(\square)=M_P^2\frac{1+\frac{2}{3}e^{\sigma_1-\sigma}\frac{\square-M^2}{2\square}(e^{\sigma}-1)}{6\square-2(\square-M^2)e^{\sigma_1-\sigma}}\, .
\end{equation}
In order to alleviate the question whether the operator above is at all invertible, we can demand that no poles exist by imposing 
\begin{equation}
e^{\sigma_1(\square)-\sigma(\square)}=3\, .
\end{equation}
This in a sense means that two entire functions are equal up to the $\log(3)$ constant term.
It leads to the following form for the function $\Fc_R(\square)$,
\begin{equation}
\label{FF}
    \Fc_R(\square)=\frac{M_P^2}{6 M^2\square}\left[(\square-M^2)e^{\sigma(\square)}+M^2\right]\, .
\end{equation}
We assume that $\sigma(0)=0$ to get a regular IR limit for $\Fc_R(\square)$. Moreover, taking $\sigma=0$ here we return to $\Fc_R=M_P^2/(6M^2)$ which is the original local Starobinsky model \cite{Starobinsky:1980te}.\footnote{$\sigma=0$ however leads to the disappearance of the Weyl squared term. This is in a sense logic because our connection of form-factors is based on the absence of ghosts paradigm. For example, we are not expecting to restore the Stelle gravity here in any low energy limit as it contains ghosts.}

We observe at this moment that the latter formula is essentially (3.18) from \cite{Koshelev:2017tvv} with one huge step forward in the understanding of the model structure: entire functions which define the two form-factors are essentially the same. This observation is made possible here thanks to using the ADM formalism which allows easier separation of physical spin-2 and spin-0 modes.

As it was shown in \cite{Biswas:2005qr}, the model (\ref{model}) can feature FLRW type solutions of the form $\square R=r_1R$ with a constant $r_1$ as long as $\Fc_R'(r_1)=0$. In \cite{Koshelev:2017tvv} it was further pointed out that one need to have $r_1=M^2$ and as such
$\Fc_R'(M^2)=0$
that for \eqref{FF} translates into 
\begin{equation}\sigma(M^2)=0.
\end{equation}
Thus, the Taylor series for an appropriate function $\sigma$ should be of the form
\begin{equation}
    \sigma(\square)=\frac{\alpha_1\square(\square-M^2)}{\Lambda^4}+(\square-M^2)\left(\frac{\alpha_2\square^2}{\Lambda^6}+\frac{\alpha_3\square^3}{\Lambda^8}+...\right).
\end{equation}
Here we set the notation $\Lambda$ for the non-locality scale and the coefficients $\alpha_n$ are some numbers of order unity. Let us keep only the leading term in the expansion for low momenta and stick to the limit $M\ll\Lambda$ (the full hierarchy would be $M\ll\Lambda\lesssim M_P$ \cite{Koshelev:2016xqb}). Then, one can expand the function $\Fc_W$ as
\begin{equation}
\label{Fw}
    \Fc_W\approx\alpha_1 M_P^2\frac{\square-M^2}{2\Lambda^4}\, .
\end{equation}
One can see that the leading term in the expansion is of the order of $M_P^2 M^2/\Lambda^4$ which leads to the Weyl squared term in the action suppressed compared to the situation when one assumes the Weyl ghost of a mass of order of the non-locality scale.

\section{Graviton production}

We are aimed at estimating the scalaron into 2-graviton vertex upon computing the triple variation of the action similar to what was done in \cite{Zeldovich:1977vgo} for non-inflationary models in GR\footnote{Note, however, that the formula for graviton creation obtained in that paper is not applicable for modified local or non-local $R^2$ gravity.}. Thus, we get post-inflationary generation of high-frequency gravitons with the initial energy $E=M/2$ in our case that much exceeds the Hubble factor $H=\frac{d}{dt}\ln a(t)$ at this epoch. The effect occurs due to small oscillations of the scale factor $a(t)$ with the high frequency $M$ during the post-inflationary stage (which is dominated by scalarons at rest), but before complete thermalization\footnote{Thus, this effect should be distinguished from creation of low-energy gravitons with the characteristic energy $E\sim H$ by a power-law FLRW expansion~\cite{Grishchuk:1974ny}, or during inflation~\cite{Starobinsky:1979ty}. In those cases, exact evolution of the backgound has to be taken into account non-perturbatively.}. Because of that, we can consider this process assuming the Minkowski approximation for the background meaning that all curvature tensors are zero and all derivatives are approximately simple partial derivatives and, most importantly, they do commute (and, actually, all connections are zero). This simplifies the analysis drastically. Moreover since we are computing the vertex, all its external legs are on shell. This implies that if, for instance, we have $\Box h_{\mu\nu}^{TT}$ factor, the corresponding expression is zero as the graviton is massless on-shell. Analogously, $\Box \varphi$ can be replaced by $M^2\phi$ for the scalaron.

Without the loss of all the features of the AID gravity, one can generalize to $\Fc_W=\Fc_W(\Box, R)$ or to assign any even more general analytic at zero dependence on the curvature tensors to the $\Fc_W$ form-factor. As long as the Minkowski limit is approached, such corrections vanish leaving us with the initial model. Thus, the Minkowski graviton propagator is not affected.
This however seems to be incompatible to do in the $\Fc_R$ form-factor. A heavy fine-tuning is required as this will lead to much more complicate equations of motion. At least, one can easily loose the desired particular Starobinsky inflationary solution.

Some routine is on the way in varying the action. The Enstein-Hilbert and $R^2$ terms do not produce a vertex corresponding to the scalaron decay to a pair of gravitons \cite{Starobinsky:1981zc}. The same is valid in the more general cases of scalar-tensor and $F(R)$ gravity \cite{Ema:2015dka,Ema:2016hlw}.
In the quadratic in $R$ term, we notice that $\delta R$ can only result in a scalaron and each $R$ must be varied at least once as it is zero on the background. Then we only can get our vertex in question from  $\sqrt{-g}\delta R\Fc_R(\Box)\delta^2R$ that again contains no graviton. Even if we include the triple in $R$ term despite the odds mentioned above, it will only result in a $\sim\varphi^3$ vertex and nothing more.

Coming to the terms quadratic in $W$, one knows that the scalaron field is coupled to gravitons through all non-conformal terms in the action. Since the local Weyl squared term is Weyl-invariant, it would not contribute to the graviton production. So, this contribution starts from the terms which have 6 derivatives acting on metric.
One can compute
\begin{equation}
    \int d^4x \delta(\sqrt{-g}WF_W(\Box)W)=\int d^4x\sqrt{-g}\sum {f_{W}}_n\sum_{l=0}^{n-1}(\Box^l\delta W)\delta\Box(\Box^{n-l-1}\delta W)+\dots
\end{equation}
where $\dots$ contain all terms which are not relevant for the scalaron decaying into 2 gravitons.
We recall that the on-shell condition assures that the $\delta W$ factors is annihilated by the $\Box$ operator if one is present.
As such we claim that all the terms with more than one d'Alembertian vanish.

Adding arbitrary cubic in curvature terms for curiosity, we make use of a powerful result by \cite{Barvinsky:1990up,Barvinsky:1994hw} that any cubic in curvature term can be written as $\Oc_1(\Box)\Rc\Oc_2(\Box)\Rc\Oc_3(\Box)\Rc$ where $\Oc_i$ are any operators and $\Rc$ are any curvature tensors, and crucially any contracted single covariant derivatives can be eliminated by adding total derivatives. Moreover, around Minkowski and for the three-vertex, each of the curvature tensors must be varied once. Triple in $R$ terms were already mentioned just above. $\delta R_{\mu\nu}$ does produce only a scalaron on-shell. As such, triple in the Ricci tensor term is out of question. $\delta W$ produces only a graviton, and thus triple in $W$ term is out of question, too.  We thus may have either $\Oc_1(\Box)R W^2$, or $\Oc_2(\Box)R_{\mu\nu}W^2$, where appropriate index contractions are presumed and no d'Alembertians act on the Weyl tensor (in order not to annihilate its variation on-shell).

We thus effectively come to only two terms which would contribute to the scalaron decay into gravitons
\begin{equation}
    L= A\, W_{\alpha\beta\gamma\delta}\,\square W^{\alpha\beta\gamma\delta}+B\, R\, W_{\alpha\beta\gamma\delta} W^{\alpha\beta\gamma\delta}
\end{equation}
with arbitrary constants $A$ and $B$ at this stage. However, $A$ can be constrained by using (\ref{Fw}), while $B$ is of order of $A$. The latter estimate follows from the fact that the graviton propagator in such models  in the de Sitter background contains $\Box-\frac23R$ combination as the argument of $\Fc_W$ \cite{Koshelev:2016xqb}. The model stability upon departure from Minkowski background requires the latter combination to be not exceeding  the order of $A$.

The corresponding decay rate is
\begin{equation}
    \Gamma=\frac{6}{\pi}\frac{M^{11}}{M_P^6} (A+2B)^2\,.
\end{equation}
Here we have followed the approach presented in \cite{Zeldovich:1977vgo}. One can readily test the validity of the above formula by dimensional arguments.
If we accommodate \eqref{Fw}, the result for $B=0$ becomes 
\begin{equation}
\label{gamma}
\Gamma_{GW}=\frac{3}{2\pi}\alpha_1^2    \frac{M^3}{M_P^2}\left(\frac{M}{\Lambda}\right)^8
\end{equation}
where we see that the established connection of form-factors dictates the scaling of the coefficient in front of $W\square W$ term with non-locality scale. 

The $W^2$ term does not produce the tree-level decay of the scalaron to gravitons but one-loop effect connected with conformal anomaly could be also relevant. The computation presented in \cite{Starobinsky:1981zc} addresses this effect and predicts the scalaron decay width,
\begin{equation}
    \Gamma_{anom}=\frac{k_1^2}{24 (2880\pi^2)^2}\frac{M^3}{M_P^2}\left(\frac{M}{M_P}\right)^4.
\end{equation}
Here $k_1$ is a coefficient in the anomalous trace of the energy-momentum tensor ($T_{\mu}^{\mu}=k_1 W^2/(2800\pi^2)+\dots$) which depends on the matter field content of the underlying theory which affects the renormalization of the coefficient in front of the $W^2$ term (typically, $k_1\sim 10$). Due to the large numerical suppression factor this decay channel is less efficient than \eqref{gamma} for $\Lambda\lesssim 10^4 M$ if $\alpha_1\sim 1$. 

As the scalaron decay rate into the Standard Model Higgs bosons minimally coupled to gravity\footnote{See \cite{Starobinsky:1981vz,Gorbunov:2012ns} for the general case of non-minimally coupled Higgs field.} is \cite{Starobinsky:1981vz,Gorbunov:2012ij} 

\begin{equation}
\label{gammaH}
    \Gamma_H=\frac{M^3}{192\pi M_P^2}\, ,
\end{equation}
the energy density of gravitons produced at the moment of reheating is smaller than the radiation energy density by the factor $M^8/\Lambda^8$,
\begin{equation}
    \frac{\Gamma_{GW}}{\Gamma_H}=288\alpha_1\frac{M^8}{\Lambda^8}\, .
\end{equation}
The produced gravitons constitute an additional component of radiation which would contribute to the number of effective relativistic degrees of freedom that is a potentially observable quantity. It is usually represented as an effective number of extra neutrino species \footnote{This equation is valid if one assumes that the Standard Model is valid until the scale of inflation.},
\begin{equation}
    \Delta N_{eff}=2.85 \frac{\rho_{GW}}{\rho_{SM}}=2.85\frac{\Gamma_{GW}}{\Gamma_H}=821\alpha_1^2\frac{M^8}{\Lambda^8}\, .
\end{equation}
Here the energy densities are taken at the moment of reheating. Taking the current Planck constraint $\Delta N_{eff}\lesssim 0.2$ \cite{Planck:2018jri} and assuming $\alpha_1\sim 1$, we obtain $\Lambda>2.8 M$ that is satisfied since it should be $\Lambda\gg M$ in order to have inflation. The conclusion here is that the overproduction of gravitons cannot happen in any higher derivative Starobinsky-like model of inflation at the perturbative level.

\section{Gravitational wave background from the scalaron decay.}

Besides the total energy density of gravitons, it is of interest to study the spectrum of produced gravitational waves. In the approximation that the gravitons are created in the process of decay of the scalaron condensate with some decay width $\Gamma_{GW}$, one can obtain a spectrum of gravitons. Following the approach of \cite{Ema:2021fdz}, one can derive for the matter-dominated stage after inflation,
\begin{equation}
\label{GWs}
    \frac{d\Omega_{GW}}{d \log{E}}=\frac{16 E^4}{M^4}\frac{\rho_{reh}}{\rho_0}\frac{\Gamma_{GW}}{H_{reh}}\frac{1}{\gamma(E)}e^{-\gamma(E)}.
\end{equation}
Here
\begin{equation}
    \gamma(E)=\left(\left(\frac{g_{reh}}{g_0}\right)^{1/3}\frac{T_{reh}}{T_0}\frac{2 E}{M}\right)^{3/2}.
\end{equation}
The quantities with the subscript 'reh' are related to the moment of reheating while the subscript '0' is for the present time. The parameters $\rho$ and $T$ stands for the total energy density and temperature, respectively, $g_0=2+7/8*2*3*4/11$ is the effective number of the Standard Model relativistic species (we assume that there are no other particles besides the Standard Model content which implies $g_{reh}=106.25$). The energy density at the reheating can be expressed through the reheating temperature,
\begin{equation}
    \rho_{reh}=g_{reh}\frac{\pi^2}{30}T_{reh}^4
\end{equation}
The quantity $H_{reh}$ is the Hubble scale at reheating which can be related to the total decay width of the inflaton $H_{reh}\sim \Gamma_{tot}$ which is expected to be dominated by the decay width to the Higgs bosons $\Gamma_H$, see \eqref{gammaH}. 

The spectrum features an exponential cutoff at high energies and a power law behaviour at lower frequencies, $d\Omega_{GW}/(d E)\propto E^{5/2}$ that corresponds to the graviton occupation number $n(E)\propto E^{-1/2}$ in this region. This specific power arises as a consequence of the fact that between inflation and reheating the Universe expansion mimics the matter-dominated stage that leads to the redshifting of the gravitational waves created at the earlier stages.

\begin{figure}[htb]
\label{GWplot}
    \begin{center}
  \includegraphics[scale=.7]{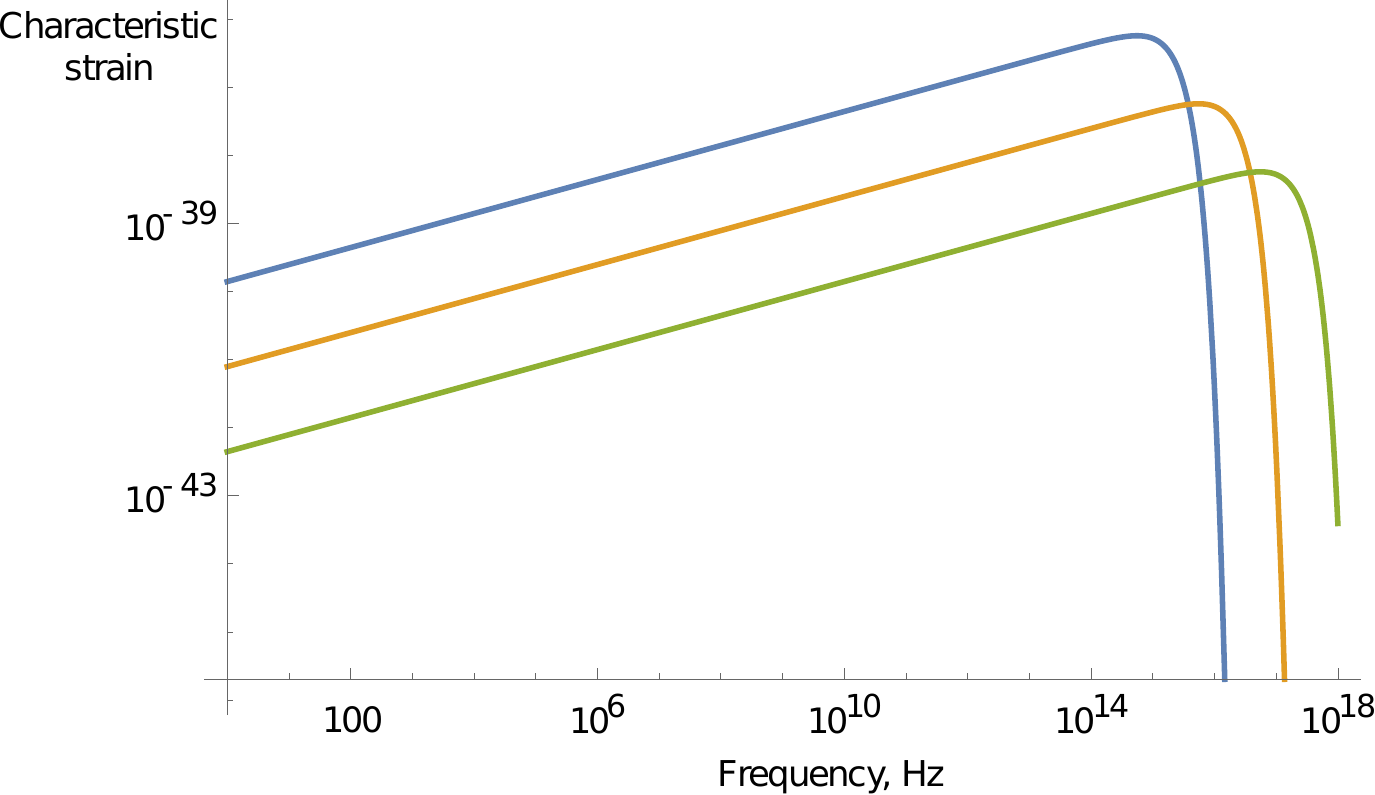}
\caption{The blue curve shows the gravitational wave signal for $T_{reh}=10^{10}$ GeV. The orange curve is for $T_{reh}=10^9$ GeV and the green curve is plotted for $T_{reh}=10^{8}$ GeV. In all cases we assume $\Gamma_{GW}/\Gamma_{SM}=10^{-3}$. The low-frequency slopes of the plots correspond to the universal $h_c\propto f^{1/4}$ behaviour following from \eqref{GWs}. The lowest characteristic strain available for future gravitational wave detectors is $10^{-24}$ for the frequencies $1-10^6$ Hz. One can see that the predicted signal is well below that level even for more intensive reheating and GW production.}
\end{center}
\end{figure}

The commonly used quantity for plotting the gravitational wave spectra is the characteristic strain $h_c(f)$ \cite{Moore:2014lga} where $f=E/(2\pi)$. It is connected to the energy density spectrum as follows
\begin{equation}
    h_c(f)=\sqrt{\frac{3 H_0^2 }{\pi f^2}\frac{d\Omega_{GW}}{d f}}.
\end{equation}
Figure \ref{GWplot} shows the gravitational wave spectrum in terms of the characteristic strain evaluated for different reheating temperatures and relations between the decay widths to gravitons and total decay width. The level of the signal is very low, compared to the projected abilities of future detectors which are at the level of $10^{-24}$. The reason for that is connected to the fact that the produced gravitational waves have very high frequences. So, the total energy density is dominated by the signal at high momenta which cannot be reached by the existing and future detectors. At the same time, the total energy density can be a significant contribution to the dark radiation and effective number of relativistic degrees of freedom if the non-locality scale is close to the scale of inflation. 

\section{Conclusions and discussion}

In this paper we addressed the question about the graviton production in a non-local ghost-free extension of the Starobinsky model. The presence of higher derivatives often leads to instabilities and enhanced effects of particle production. For this reason, the overproduction of the gravitational waves after inflation could be a serious issue appearing in the Starobinsky model embedded to a UV complete gravity by means of adding infinite number of derivatives in a propagator. However, we have shown that this would never happen in a perturbative decay of the scalaron to gravitons taking place after inflation. We computed the gravitational wave signal, as well as the contribution of graviton radiation to the number of effective relativistic degrees of freedom bounded by BBN and CMB. We found that, given that the gravitational waves have high frequencies, negligible signal for the detectors can be at the same time visible as dark radiation. Thus, in the class of models under consideration, the strongest constraint on the scale of suppression of the higher derivative operators would come from CMB. The gravitational wave signal would lie in the very high frequency domain, being invisible even for the future detectors.

This conclusion will be applicable to any model of inflation allowing for the process of the inflaton decay to gravitons. Typically, the lower is the reheating temperature, the stronger is the dark radiation constraint on the scale of quantum gravity effects. However, the effect of the perturbative graviton production through the inflaton decay would never result in significant gravitational wave signal.
In the present consideration we do not consider additional
(quasi)-particles in the model which might arise from the complex conjugate
poles arising once one departs from the background solution. Even would such particles be considered, their effect can be contained by choosing the higher derivative form-factor as suggested in \cite{Koshelev:2020fok}. Nevertheless, in general they present another source of
gravitons.
Also one may expect the presence of non-perturbative effects, still being relevant for the graviton production. We leave both these questions for future study.

Another source of gravitational waves in both local and non-local $R^2$ inflation would be related to the effects of inhomogeneous reheating occurring after the inflaton clumping due to the evaporation of the formed structures. Estimates of the strength of the signal have shown that it could be the dominant contribution to the stochastic GW background at lower frequencies available for LIGO and LISA \cite{Jedamzik:2010hq}. However, accurate numerical computations are required for final predictions of the signal strength.





\section*{Acknowledgements}
AK is supported by FCT Portugal investigator project IF/01607/2015. AAS is supported by the RSF grant 21-12-00130. AT is supported by the European Union Horizon 2020 Research Council Grant No. 724659 MassiveCosmo ERC2016COG and by a Simons Foundation Award ID 555326 under the Simons Foundation Origins of the Universe initiative, Cosmology Beyond Einstein’s Theory. 
\section*{Dedication}
We dedicate this work to the memory of Valery Rubakov, a distinguished physicist and an unforgettable personality. In addition to his fundamental contributions to quantum field theory and cosmology, he established a well-known theoretical physics school and was able to maintain brilliant traditions of physics community in difficult circumstances. Coming from Valery's school, AT is especially indebted to him for numerous deep discussions, constant personal support and extremely fruitful criticism. Spirit of Valery will keep warming our hearts and souls.

\end{document}